# MULTIFRACTAL CHARACTARIZATION AS A FUNCTION OF TIMESCALE IN THE LIGHT CURVES WITH PLANETARY SIGNAL OBSERVED BY THE KEPLER MISSION

**Caracterização multifractal como uma função da escala de tempo em curvas de luz com sinal planetário observada pela missão Kepler**

**Fernando José Silva Lima Filho**
Undergraduate student in Physics
Institution: Federal University of Ceará
Address: Campus do Pici, Fortaleza – CE
E-mail: fernandolima@fisica.ufc.br

**Vitor Marcelo Belo Ferreira**
Undergraduate student in Physics
Institution: Federal University of Ceará
Address: Campus do Pici, Fortaleza – CE
E-mail: vitormarcelobeloferreira@alu.ufc.br

**Paulo Cleber Farias da Silva Filho**
M.A. student in Astrophysics
Institution: Federal University of Ceará
Address: Campus do Pici, Fortaleza – CE
E-mail: clebersilva@fisica.ufc.br

**Fernando Oliveira da Silva Gomes**
PhD student in Astrophysics
Institution: Federal University of Ceará
Address: Campus do Pici, Fortaleza – CE
E-mail: fernandogomes@fisica.ufc.br

**Brício Warney de Freitas Alves**
M.A. student in Astrophysics

Institution: Federal University of Ceará
Address: Campus do Pici, Fortaleza – CE
E-mail: bricio@fisica.ufc.br

Sarah Gomes Aroucha Barbosa
PhD student in Astrophysics
Institution: Federal University of Ceará
Address: Campus do Pici, Fortaleza – CE
E-mail: sarah@fisica.ufc.br

Thiago de Melo Santiago
PhD student in Astrophysics
Institution: Federal University of Ceará
Address: Campus do Pici, Fortaleza – CE
E-mail: thiago@fisica.ufc.br

Daniel Brito de Freitas
Professor Doctor
Institution: Federal University of Ceará
Address: Campus do Pici, Fortaleza – CE
E-mail: danielbrito@fisica.ufc.br

## ABSTRACT

Astrophysical data, in the domains of time, involve a wide range of stellar variability phenomena, among them the magnetic activity of the order of a few hours until the signature of an extra-solar planet which can cover a scale of time of a few days until tens of years. Numerous instruments are being developed to detect Earth-sized exoplanets. Exoplanets with this dimension challenge scientific instrumentation and the field of research in the data processing. In this context, our study offers a powerful framework to explain dynamical properties as a function of timescale in light curves with the planetary signal. For that, we selected the stellar target Kepler-30 to test our methods and procedures. In this sense, we investigate the multifractal behavior of the Kepler-30 system composed of a sun-like star with a rotation period of ~16 days and three planets with masses between 2 Earth and 2.5 Jupiter masses. Furthermore, this system has an orbital period varying from 29 to 143 days and orbits almost coplanar. This system is highly interesting because starspots dynamics are strongly affected by the passing of a planet in front of the star. We used about 1600 days of high-precision photometry collected by the Kepler mission to investigate the quasi-periodic variation caused by the rotation of the star and the effect of spot evolution as a function of timescale. We applied indexes extract from multifractal analysis to model the flux rotational modulation induced by active regions. Our results that stellar flux variations in Kepler-30 star caused by rotational modulation can be replicated in detail with just four recent-known multifractal indexes. These indexes will greatly simplify spot modelling of current TESS and future PLATO data.

## RESUMO

Os dados astrofísicos, no domínio do tempo, envolvem uma ampla gama de fenômenos de variabilidade estelar, entre eles a atividade magnética da ordem de algumas horas até a assinatura de um planeta extra-solar que pode abranger uma escala de tempo de poucos dias até dezenas de anos. Numerosos instrumentos estão sendo desenvolvidos para detectar exoplanetas do tamanho da Terra. Exoplanetas com essa dimensão desafiam a instrumentação científica e o campo de pesquisa em processamento de dados. Nesse contexto, nosso estudo oferece uma estrutura poderosa para explicar propriedades dinâmicas em função da escala de tempo em curvas de luz com sinal planetário. Para isso, selecionamos o alvo estelar Kepler-30 para testar nossos métodos e procedimentos. Nesse sentido, investigamos o comportamento multifractal do sistema Kepler-30 composto por uma estrela semelhante ao Sol com período de rotação de ~16 dias e três planetas com massas entre 2 massas terrestres e 2,5 de Júpiter. Além disso, este sistema possui período orbital variando de 29 a 143 dias e órbitas quase coplanares. Este sistema é altamente interessante porque a dinâmica das manchas estelares é fortemente afetada pela passagem de um planeta na frente da estrela. Usamos cerca de 1600 dias de fotometria de alta precisão coletada pela missão Kepler para investigar a variação quase periódica causada pela rotação da estrela e o efeito da evolução do ponto em função da escala de tempo. Aplicamos índices extraídos da análise multifractal para modelar a modulação rotacional do fluxo induzida por regiões ativas. Nossos resultados de que as variações de fluxo estelar na estrela Kepler-30 causadas pela modulação rotacional podem ser replicadas em detalhes com apenas cinco índices multifractais conhecidos recentemente. Esses índices simplificarão bastante a modelagem pontual dos dados atuais do TESS e do PLATO futuro.

**Palavras-Chave:** variabilidade estelar, missão Kepler (NASA), Kepler-30 (KOI-806) star, análise multifractal

# 1 INTRODUCTION

Stellar variability is fundamental in stellar astrophysics and plays an important role in the time evolution of magnetic indicators, such as rotation, flares, and magnetic cycles (Kraft, 1965; Skumanich, 1972; Kawaler, 1988). Furthermore, the variability may also offer valuable information on stellar magnetism, mixing in the stellar interior, tidal interactions in close binary, as well as the relationship between the rotation and magnetic activity, which has implications for the detectability of planets orbiting solar-type stars (Maxted et al., 2016; de Freitas et al., 2021). These stars comprise the vast majority of the Kepler exoplanet field targets and, in our context, they are very important to investigate the stellar variability due to magnetic activity. In particular, starspots and active regions on the stellar surface modulate the stellar brightness on the rotation timescale of the star (Walkowicz and Basri, 2013). As mentioned by Sanchis-Ojeda et al. (2012), the effects produced by the presence of starspots, such as quasi-periodic modulation caused by rotation and, shorter-term "anomalies" caused by the transit of a planet in front of a spot, affect the profile of flux.

The *Kepler* space mission, during its operation period from 2009 to 2013, found systems with several planets, among them a curious and particularly interesting system named Kepler-30 studied recently by de Freitas et al. (2021). Kepler-30 (KOI-806) star is a fast rotator with nearly solar mass and radius, and a rotation period of ~16.0 days based on a Lomb-Scargle periodogram as reported by Sanchis-Ojeda et al. (2012) and Lanza, das Chagas and De Medeiros (2014). Kepler-30 is a system composed of three planets denoted by Kepler-30 *b*, *c*, and *d* with orbital periods of ~29, 60, and 143 days, respectively Sanchis-Ojeda et al. (2012) and Panichi et al. (2018). This system is unique because the orbits of its planets are nearly aligned, a similar behavior to our Solar System and, therefore, low obliquity observed. As cited by de Freitas et al. (2021), low-obliquity systems suggest a strong correlation exists between the timing of transit and a local minimum in the quasi-periodic modulation caused by rotation. In other words, spots can be eclipsed by one or more planets. For the Kepler-30 system, planet Kepler-30*c* (13 Earth's radius) yields the most significant effect over the starspots with nearly four stellar rotation periods between the transits, i.e., ~60 days. Thus, shorter-term correlations in flux can be found when the transit will be removed, and more robust treatment using multiscale methods is necessary.

Over the last years, several methods were used for the estimation of (multi)fractal structure of time series in astrophysical databases (e.g. de Freitas et al. 2016, 2017; Belete et al. 2018; Franciscis 2019; de Freitas et al. 2019a, 2019b). The estimation of local fluctuations and long-term dependency of astrophysical time series is a problem that has been recently studied to understand the effect of the long-memory process (de Freitas et al., 2019a, 2019b). As mentioned by de Freitas et al. (2013), the global Hurst exponent (Hurst, 1951) is a powerful parameter for analyzing CoRoT and *Kepler* time series, which is mainly used to elucidate the persistence due to rotational modulation of the time series. Since 2017, a novel technique proposed by Gu and Zhou (2010), known as the multifractal detrended moving average (MFDMA) method, has been applied for the multifractal characterization of the *Kepler* light curves (de Freitas et al., 2017, 2019a, 2019b, 2021). The MFDMA method filter out the local trends of non-stationary series by subtracting the local means. MFDMA method investigates the local fluctuations of time series using an a priori (fixed) timescale. According to Wang, Shang, and Cui (2014), fixing a priori scaling ranges may lead to a crossover that falls within the scaling range by mistake, and, therefore, the results could be biased. A method to avoid mistakes due to fixed scaling ranges is to measure the multifractal properties considering the entire timescales. To this end, the MFDMA method as a function of timescale was developed in this work.

This paper is structured as follows. Observations and stellar and planetary parameters are shown in Section 2 when we also analyze the Kepler-30 light curves in two pipelines named PDC and SAP. In particular, we investigate the impact of the PDC pipeline over stellar variability when compared to SAP data, giving an overview of the characteristics of the PDC and SAP data. In the next Section, we present the MFDMA method as a function of the timescale from cadence level until values of the order of the rotation period. Then we applied our method to the Kepler-30 data in Section 4 when the results and discussions are presented. In the last section, our final remarks are summarized in detail.

# 2 KEPLER-30 LIGHT CURVES AS A FUNCTION OF TIMESCALES

The Kepler mission performed 17 observational runs, also known as quarters with a time window of ~90 days each, for ~200 000 stellar targets. In particular, the light curves of these targets are composed either of long cadence, i.e., data sampling every 29.4 min (Jenkins et al., 2010b), or short cadence with sampling every ~1 min observations (Van and Caldwell, 2009; Thompson et al., 2013). A detailed discussion of the public archives can be found in many Kepler team publications (Borucki et al., 2009, 2010; Batalha et al., 2010; Koch et al., 2010; Basri et al., 2011, 2013). Regarding data format, the Kepler mission provides two types of data named by acronyms PDC and PDC. Simple Aperture Photometry (SAP) data are processed using a standard treatment only removing more relevant spacecraft artifacts, whereas Pre-Search Data Conditioning (PDC) is processed through a refined treatment which removes more thermal and kinematics effects due to spacecraft operation (Van et al., 2010). A description in more detail can be found in the Kepler pipeline proposed by Jenkins et al. (2010a). The difference between these data is extremely relevant for our work.

Kepler-30 is a system composed of a sun-like star with a rotation period of ~16 days and three transiting planets with masses between 2 Earth and 2.5 Jupiter masses. Its orbital period varies from 29 to 143 days with orbits almost coplanar (Fabrycky et al., 2012). In addition, the mean evolutionary time of the spot pattern is ~22 days. Kepler-30 star is highly interesting because starspots dynamics are strongly affected by the passing of a planet in front of the star (Fabrycky et al., 2012). The data of Kepler-30 covers four years of high-precision photometry measurements collected by Kepler's mission to investigate the quasi-periodic variation caused by the rotation of the star and the effect of spot evolution as a function of timescale. To analyze this scenario, observations used for our work are long-cadence temporal sampling, considering that our interest is to investigate the out-of-transit variations on the timescale of stellar rotation or longer.

**Figure 1:** $\tau$-light curves generated from PDC data using the difference $S(t + \tau) - S(t)$. To make the figure cleaner, we have omitted the axes which are actually defined as time versus photometric flux.

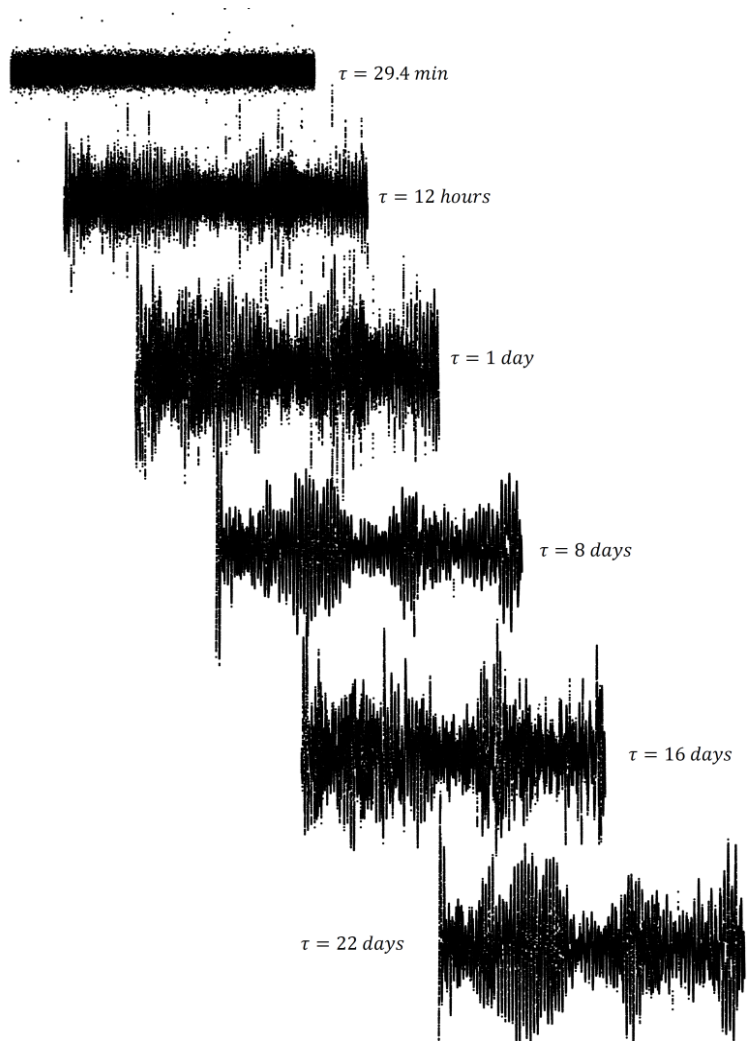

**Figure 2:** Idem Figure 2 for SAP data.

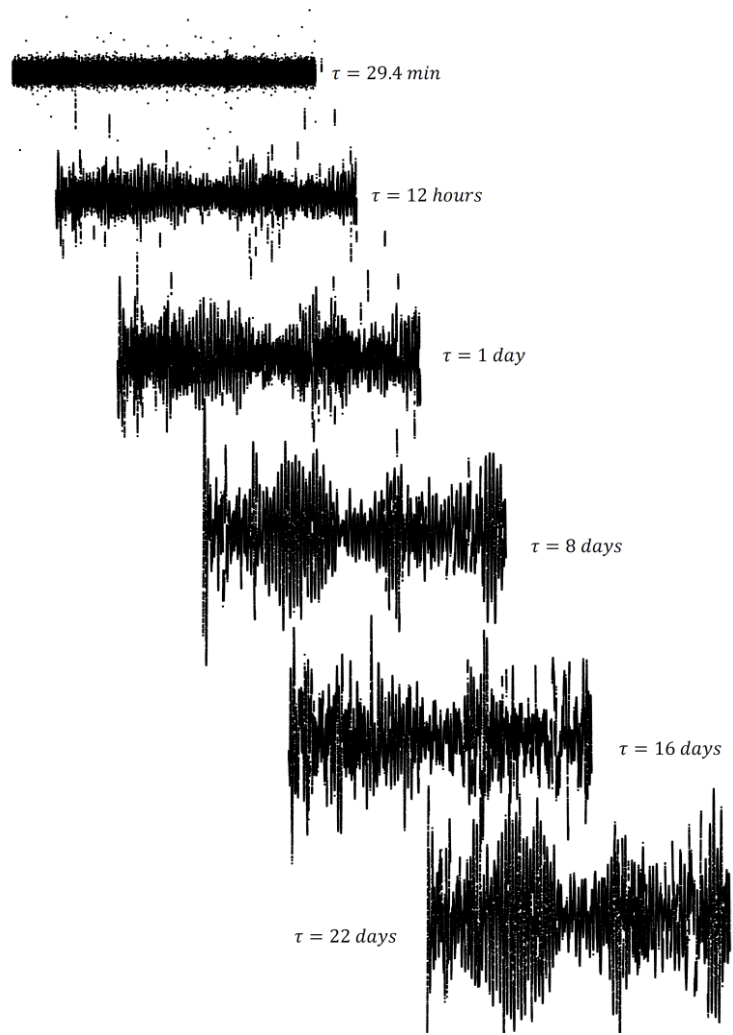

The literature points out that the PDC pipeline modifies stellar variability on timescales longer than a few days and, therefore, hampers distinguishing between astrophysical and instrumental signals. By the way, the removal of variability, even partially, from the light curves can affect the analysis, as mentioned by Gilliland et al. (2015) and de Freitas et al. (2019b). Gilliland et al. (2015) and Lanza et al. (2019) mentioned that the PDC pipeline sometimes affects the modulations on time scales longer than a few days because it has been designed to detect planetary transits. Consequently, this limitation does not preserve the intrinsic stellar variability. In special, these modulations are particularly crucial for Kepler-30 because it shows a large light curve amplitude higher than the median of the errors of the photometric measurements as given by the Kepler pipeline with typical modulation timescales of dozens of days. On the other hand, SAP data are less affected by this attenuation.

For these reasons, we decided to investigate the behavior of fluctuations in the out-of-transit flux variations of Kepler-30 star considering both pipelines. The PDC and SAP light curves used in the present work can be seen in Figures 1 and 2 detailed in de Freitas et al. (2021). Thus, we build $\tau$-light curves, fluctuations of increments $S(t + \tau) - S(t)$ due to its variability over a timescale, where $S$ denotes the data series and $\tau$ the characteristic time that defines the scale of fluctuations represented by difference $S(t + \tau) - S(t)$ as shown in Figures 1 and 2 present in this paper. In our study, $\tau$ varies from cadence level (29.4 min) until values of the order of the rotation period. This procedure has been used to describe magnetic field fluctuations due to solar wind on a large scale range as proposed by de Freitas and De Medeiros (2009).

## 3 MULTIFRACTAL ANALYSIS

Generally speaking, our methodology is based on the geometry of the multifractal spectrum, which was inspired in figure 2 developed by de Freitas et al. (2017). This spectrum is the final step of the algorithm by Gu and Zhou (2010) and consists of the analysis of small and large magnitude fluctuations, $F_q(n)$, power law of the type $F_q(n) \propto n^{h(q)}$, where $n$ is a scale factor and $h(q)$ is denoted as the generalized exponent of Hurst. From the exponent $h(q)$ the exponent of the multifractal scale defined in the form $\tau(q) = qh(q) - 1$ is extracted. Finally, through a Legendre transform, we can obtain the Holder exponent $\alpha(q) = \frac{d\tau(q)}{dq}$ and the multifractal spectrum $f(\alpha) = q\alpha - \tau(q)$ (cf. de Freitas et al. 2021). For a monofractal time series, $\tau(q)$ is a linear function given by $qH - 1$, where $H$ is the Hurst exponent global. For a multifractal signal $\tau(q)$ is nonlinear and the multifractal spectrum $f(\alpha)$ takes the form indicated by the parabolic.

We tested the set of five multifractal indicators that were extracted from the above equations proposed by de Freitas et al. (2017). In this paper, the authors used the referred figure 2 to describe the shape of the multifractal spectrum. Here, we show the detailed description of the list of five multifractal indicators as described below:

**Figure 3:** *Top panel*. Multifractal analysis for the PDC (without planetary transit) light curve of the Kepler-30 star in three versions: original (red), shuffled (green) and phase-randomized surrogated time series (blue). *Left middle panel*. Multifractal fluctuation function $F_n(q)$ obtained from PDC light curve of the Kepler-30 developed by de Freitas et al. (2021). Each curve corresponds to different fixed values of $q = -5, -4, \ldots, 4, 5$. *Right middle panel*. Holder exponent $h(q)$ as a function of parameter $q$. *Left bottom panel*. Comparison of the multifractal scaling exponent $\tau(q)$ for three versions. *Right bottom panel*. Multifractal spectrum $f(\alpha)$ of three time series.

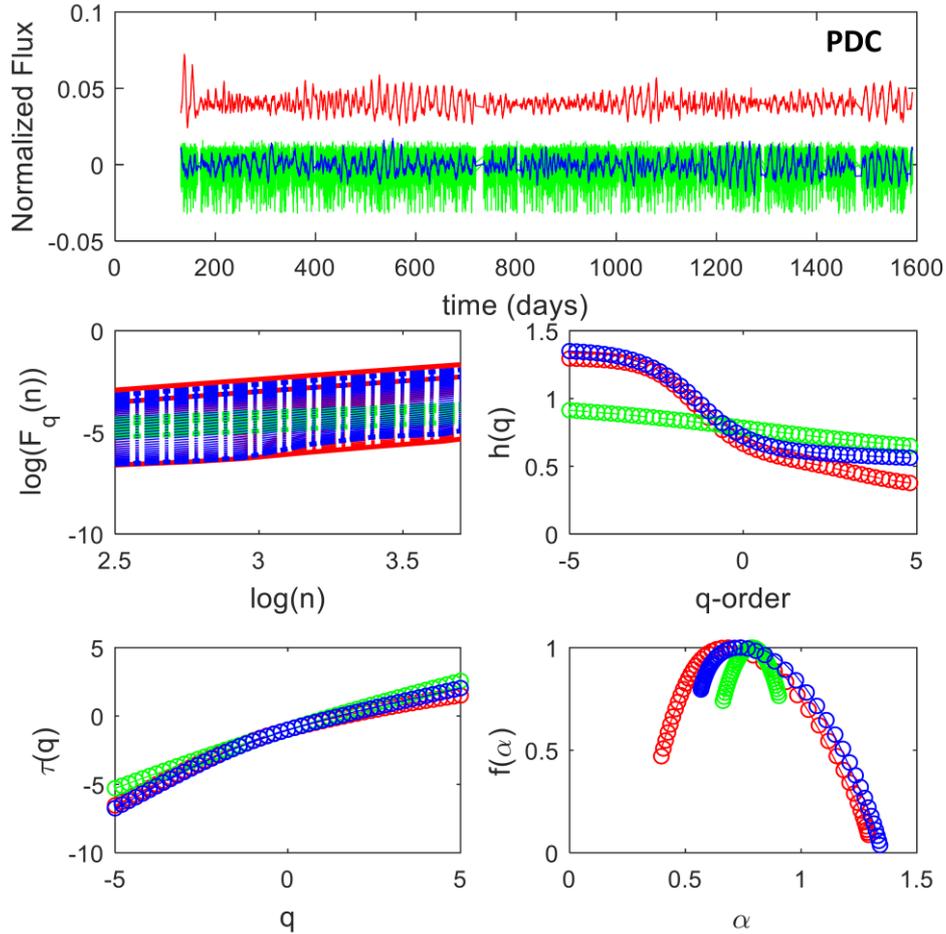

### The parameter $\alpha_0$

The $\alpha_0$ parameter delivers valuable information about the stellar variability, where its value is associated with the maximum of the multifractal spectrum f(α). As an example, higher values indicate that $\alpha_0$ is less correlated and has a fine structure (Krzyszczak et al., 2018; de Freitas et al. 2021). In addition, this parameter is strongly affected by signal fluctuations as cited by de Freitas et al. (2021).

### Singularity parameter $\Delta f_{\min}(\alpha)$

Parameter $\Delta f_{\min}(\alpha)$ characterizes the broadness, which is defined as the difference $f(\alpha_{\max}) - f(\alpha_{\min})$ of the singularity spectrum defined by $f(\alpha)$. This difference provides an estimate of the spread in changes in fractal patterns. Considering that $\Delta f_{\min}(\alpha)$ denotes the frequency ratio of the smallest ($q < 0$) to the largest ($q > 0$) fluctuations, $\Delta f_{\min}(\alpha) > 0$ means that the smallest fluctuations are more frequent than largest fluctuations. As quoted by Ihlen (2012), the largest fluctuations imply that the singularities are stronger, whereas the smallest fluctuations indicate that the singularities are weaker (Tanna and Pathak, 2014).

**Figure 4:** Same procedure proposed in Figure 3 for the SAP light curves (without planetary transit).

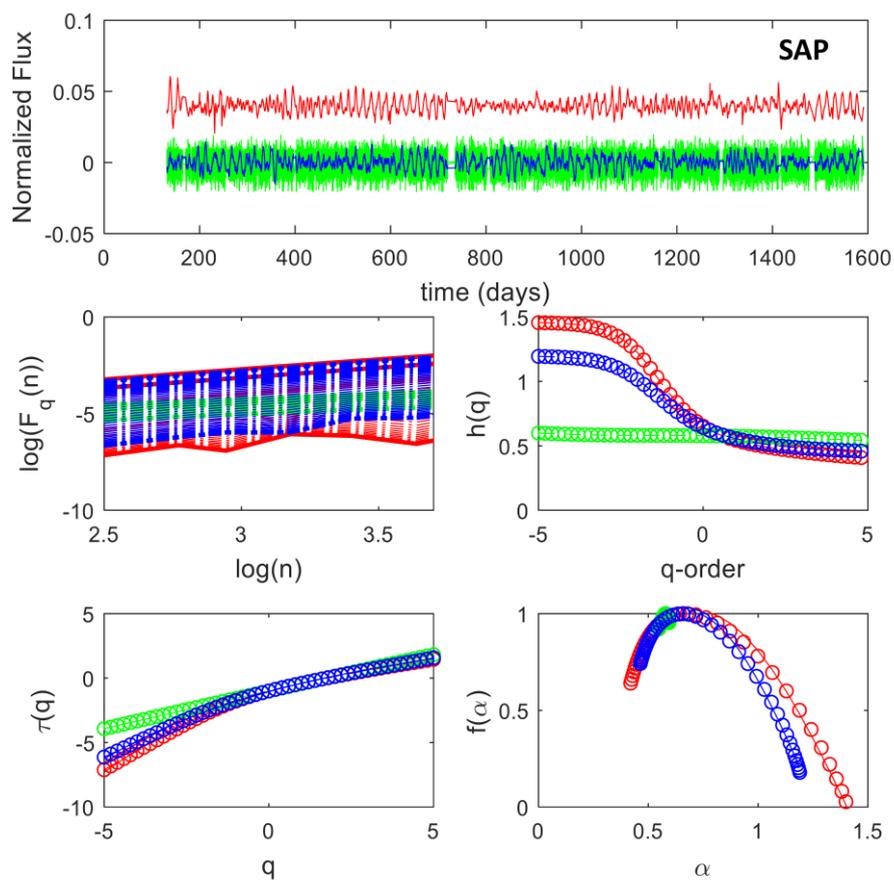

### Degree of asymmetry (*A*)

This index, which also called the skewness in the shape of the $f(\alpha)$ spectrum, is expressed as the following ratio $A = \frac{\alpha_{\max} - \alpha_0}{\alpha_0 - \alpha_{\min}}$, where $\alpha_0$ is the value of $\alpha$ when $f(\alpha)$ is maximal. The value of this index $A$ indicates one of three shapes: right-skewed ($A > 1$), left-skewed ($0 < A < 1$) or symmetric ($A = 1$). The left endpoint $\alpha_{\min}$ and the right endpoint $α_{max}$ represent the maximum and minimum values of the singularity exponent, respectively.

**Degree of multifractality ($\Delta\alpha$)**

This index represents the broadness given by relationship $\Delta\alpha = \alpha_{max} - \alpha_{min}$, where $α_{max}$ and $α_{min}$ are as defined above. A low value of $\Delta\alpha$ indicates that the time series is close to fractal, and the multifractal strength is higher when $\Delta\alpha$ increases (de Freitas et al., 2009, 2017). The bigger the width of the multifractal spectrum $\Delta\alpha$, the more heterogeneous the fractal scaling properties of a time series. According to Makowiec and Fuliaski (2010), if $\Delta\alpha$ is lesser than 0.05, then the monofractal behavior of the spectrum should be assumed.

**Figure 5:** Multifractal indicadors calculated for original PDC and SAP light curve based on the equations shown in Section **Multifractal Analysis** as a function of timescale $\tau$, as well as $\tau$-light curves shown in Figures 1 and 2.

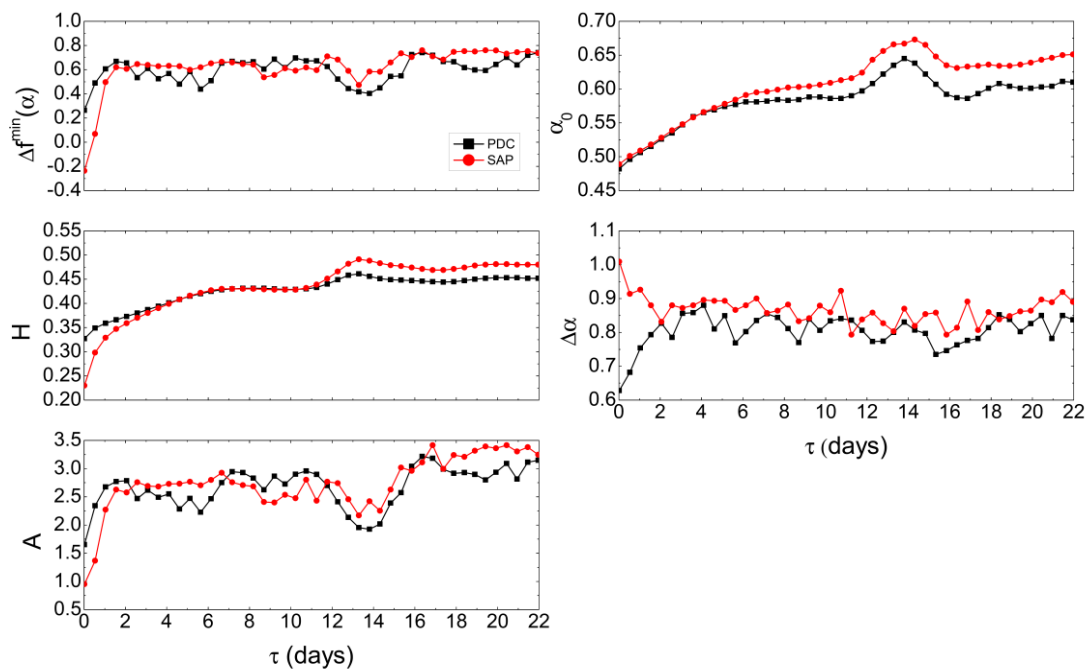

**Hurst exponent ($H$)**

According to de Freitas et al. (2017), the Hurst exponent $H$ can be defined by the second-order statistical moment ($q = 2$) extracted from multifractal formalism (Ihlen, 2012). In our study, $H$ is defined as specific Holder exponent $h(2)$. De Freitas et al. (2013) mentioned that the exponent $H$ denotes Brownian motion when $H = 1/2$, i.e., past and future fluctuations are uncorrelated. On the other hand, if $H > 1/2$, fluctuations are linked to the long-term persistence signature, i.e., an increase in values will most likely be followed by another increase in the short term, and a decrease in values will most likely be followed by another decrease in the short term. Finally, $H < 1/2$ indicates the fluctuations tend not to continue in the same direction, but instead turn back on themselves, which results in a less-smooth time series (Hampson & Mallen, 2011).

In particular, the pink noise $H = 1$ separates between the noises $H < 1$ that have more apparent fast evolving fluctuations and random walks $H > 1$ that have more apparent slow evolving fluctuations (Ihlen, 2012). This issue is relevant when we analyze the time-dependence of $H$ (see Section Results and Discussions). Since 2013, de Freitas et al. (2013) have shown that the scaling exponent a.k.a Hurst exponent is one of the most relevant (multi)fractal parameters when it comes to time series with rotational modulation.

In this context, we also investigate the behavior of these indicators based on a new approach that allows us to extend the analysis of stellar variability to include the dependence on the timescale, here defined as $\tau$ and detailed above. This procedure considers the dynamical effects caused by autocorrelated light curves as illustrated in Figures 1 and 2. With it, we can investigate separately the impacts of noise, non-stationarity, and rotational modulation. Consequently, we stipulated that the $\tau$-values should vary between the data cadence (~30 min, the noise level) to values slightly higher than the 16-day rotation period of the Kepler-30 star (see Figures 1 and 2).

**Origin of multifractality**

Basically, there are two types of time series, namely, shuffling and phase-randomized surrogated time series, which are used to verify the different origins of the multifractality. In particular, the shuffling series destroys the memory but preserves the distribution of the data with $h(q) = 1/2$. In this case, the source of the multifractality in the time series only presents long-range correlations (de Freitas et al., 2017). On the other hand, the origin of multifractality can also be due to the presence of non-linearity in the original time series. General speaking, the non-linear effects can be weakened by creating phase-randomized surrogates, thereby preserving the amplitudes of the Fourier transform and the linear properties of the original series by randomizing the Fourier phases (Norouzzadeha, Dullaertc and Rahmani, 2007; de Freitas et al., 2017). In this case, if the origin of multifractality is non-linearity that is obtained by the phase randomized method, the values of $h(q)$ will be independent of q, and $h(q) = 1/2$ will not necessarily hold.

## 4 RESULTS AND DISCUSSIONS

As mentioned in the final of the previous section, multifractality can be due to either long-term correlations or a fat-tailed probability distribution due to non-linearity. In the case of the Kepler light curves, it was already shown that the first source of multifractality applies and that the second one has no effect (de Freitas et al., 2017, de Freitas et al. 2021). The top panel from Figures 3 and 4 shows the average results of 200 realizations of the shuffled and phase-randomized surrogates of PDC and SAP light curves in green and blue colors, respectively. In fact, as can see in these figures, the shuffled procedure destroyed the correlations, i.e., the Holder exponent is slightly flat in $h(q) \sim 0.65$ for PDC data, whereas it is flat for SAP data, resulting closely from the value for white noise ($h(q) \sim 1/2$). However, the Holder exponent of phase-randomized series is similar to the original series, mainly, for values of $q > 0$. These findings suggest that the multifractality of rotational modulation is due to both long-range correlation and non-linearity, but the source of multifractality is mainly long-range correlation, which is consistent with the results of MFDMA method. Similar results are obtained when we use $\tau(q)$ as a reference.

**Table 1:** The table includes five multifractal indicators extracted from the standard MFDMA method for PDC and SAP light curves distributed in the Original ($O$), Shuffled ($S$), and Phase-randomized ($P$) versions.

|     | $\Delta f_{\min}(\alpha)$ | $\alpha_0^O$ | $\alpha_0^S$ | $\alpha_0^P$ | $H_O$ | $H_S$ | $H_P$ | $\Delta\alpha_O$ | $\Delta\alpha_S$ | $\Delta\alpha_P$ | $A_O$ | $A_S$ | $A_P$ |
| --- | --- | --- | --- | --- | --- | --- | --- | --- | --- | --- | --- | --- | --- |
| PDC | 0.58 | 0.58 | 0.55 | 0.59 | 0.42 | 0.51 | 0.43 | 0.98 | 0.11 | 0.86 | 2.99 | 1.15 | 3.34 |
| SAP | 0.67 | 0.64 | 0.52 | 0.65 | 0.48 | 0.53 | 0.49 | 0.91 | 0.05 | 0.94 | 2.89 | 1.82 | 2.68 |

We calculate the MFDMA fluctuation functions $F_n(q)$ as a function of window size n for two light curves named by PDC and SAP. The scale parameter n is varied from 10 to $N/10$, where $N$ is the total number of data points, and the parameter q is varied from -5 to 5 in steps of 0.2. The scaling pattern of $F_n(q)$ of original data (red lines) is shown in the second panel of Figures 3 and 4. We also repeat the analysis for a set of 200 randomly shuffled series as well as 200 phase-randomized series (blue and green lines, respectively).

According to Figures 3 and 4, the multifractality due to correlation is stronger than that due to nonlinearity as indicated by the positive values of $\Delta f_{\min}(\alpha)$ in both PDC and SAP data. This value does not imply that there are only long-term correlations in fluctuations, but that non-linearities are very weak. As a result, a time-independent structure with $h(q) \sim 1/2$ is shown in the shuffling time series (see Figs. 3 and 4). After shuffling, we verified that all light curves exhibit a smaller degree of multifractality $\Delta\alpha$ than the original one. This result can be emphasized by the multifractal spectra (green curves from right top panels of figures 3 and 4). In general, this analysis is not conclusive for explaining the source of this behavior. In this case, it is necessary to verify the behavior of $\tau(q)$. As illustrated by Figures 3 and 4 (left bottom panels), there is a strong dependence of shuffled and original $h(q)$'s on $q$, which is clearest for $q < 0$, where a deeper right tail occurs in multifractal spectra (see right bottom panels from Figs. 3 and 4). In contrast, the dependence on q can be roughly neglected for the randomized phase data in both PDC and SAP data. Thereby, the two types of multifractality appear in $q < 0$, a domain where the rotational modulation occurs. Basically, the correlations and non-linearities are negligible for $q > 0$, where strong fluctuations due to the noise appear.

Because the presence of rotational modulation and noise can affect the geometric indicators to a given degree, we decided to investigate the changes in the geometrical profile of the multifractal spectrum by comparing the indicators calculated from the original series with those obtained from the surrogate series. Consequently, it is possible to find the source(s) that affect the values of five multifractal indicators following the behavior of these indicators when using the surrogate time series.

Firstly, the relation between the values of $\alpha_0$ in PDC and SAP light curves shows a clear difference between them. It is interesting to note that SAP data has larger $\alpha_0$ (for original data) values than the PDC light curve, which means that in this data, the fluctuations governing the rotational modulation are more correlated and have a less fine structure (a structure more regular in appearance) compared to the fluctuations governing the SAP data.

Secondly, as suggested by de Freitas et al. (2017), the values of H for the surrogate data indicate that the shuffling data destroys the correlations and, therefore, $H$ tends to 0.5, whereas the phase-randomized data recover a value very close to those found for the original series, meaning that the non-linearities have little affect rotational modulation even if it is aperiodic.

Contrastingly, the degree of multifractality $\Delta\alpha$ changes more vastly among the surrogate and original light curves. Because this parameter is connected to the richness of the data structure, we highlight the $\Delta\alpha$ for both original light curves indicate such data may promote some values of the fluctuations, making the signal structure greater rich. Nevertheless, there is an important detail. Considering the original light curve, as $\Delta f_{\min}(\alpha)$ is the smallest value, the broadness of $\Delta\alpha$ is mixing between strong and weak fluctuations and, therefore, the noise has an important effect on the data, as can be emphasized by the small values of $\Delta\alpha$ for the shuffled light curve. Although, following the criterion of Makowiec and Fuliaski (2010), the shuffling process does not completely reduce the series to a monofractal.

It is possible to observe the similarity of asymmetry parameter A among the light curves for both the original and the randomized phase data, indicating that the non-linearity effects are weaker than the long-trend correlations. As shown in Figures 3 and 4, it is interesting that generally, the multifractal spectra $f(\alpha)$ are rather right-skewed, suggesting that fine structures are more frequent. However, the extreme events become slightly stronger for the SAP data, as the left side of the spectrum of this series is slightly deeper than that from PDC data. The result of the standard multifractal approach for these five multifractal indicators segregated by three different versions is presented in Table 1.

Now, this scenario can change when the effects of timescale are considered. For PDC and SAP from Kepler-30, the standard multifractal characterization was performed following the procedures proposed by de Freitas et al. (2017), and also for the series obtained from surrogates of the original data sets. For all of the data series, five multifractal indicators were obtained and their values as a function of timescale in agreement with τ-light curves shown in Figs. 1 and 2. Figure 5 highlights the behavior of these five indicators as a function of τ measured in days.

Considering Figure 5, when τ changes from small to large values, the behavior of $\alpha_0$ and $H$ show a rapid rise to their peak values at ~14 days, then become more or less constant showing a plateau followed by a slightly minimum peak at ~16 days (rotation period). However, according to indicators $\Delta f_{\min}(\alpha)$, $\Delta\alpha$, and $A$, there are oscillations that remain virtually constant as $\tau$ increases. In addition, there is also a stronger jump when $\tau$ is small in both PDC and SAP light curves. This behavior may be the result of the amplified impact of small fluctuations that in turn depends on several effects such as the evolution of active regions and the differential rotation (cf. de Freitas et al. 2021).

In conclusion, it is interesting to address a comparison of the parameters of the multifractal spectra between the two data, namely PDC and SAP. It allows us to address the question of whether the MFDMA method can be applied as an indicator of the changes in the dynamics of fluctuations as a function of the *Kepler* pipeline. It can be observed that all of the parameters change only slightly and the average behavior is the same for both PDC and SAP pipelines. On the other hand, the degree of multifractality $\Delta\alpha$ as a function of timescale is more developed in the SAP data, whereas the asymmetry $A$ for SAP time series is slightly more positive. Some differences between both time series can be also seen when analyzing the absolute differences of Hurst exponents for the original and shuffled data or original and surrogates. Even though we do not see a change between both time series, it is noticed that the source of multifractality due to the contribution of long-range correlations is dominant.

## 5 FINAL REMARKS

In this study, we analyze the evolution of five multifractal indicators of the Kepler light curves of the moderately young Sun-like Kepler-30 star accompanied by a three-planet system, in both PDC and SAP pipelines using an MFDMA-based multifractality analysis approach in standard and timescale versions. In addition, PDC and SAP light curves were examined using the dependence of five indicators on the timescale $\tau$. We concentrate not only on the fact that the rotational modulation has multifractal properties, as already mentioned by de Freitas et al., (2013b, 2016, 2017, 2019a, 2019b, 2021) but also on some studies on these properties as a function of timescale.

The results shown in Figures 3 and 4 highlight that five multifractal indicators of Kepler-30 data have a strong relationship with the timescale $\tau$ and, therefore, indicating the results of the standard MFDMA method using a fixed timescale are limited. Then, we systematically investigate the dynamic behaviors of the small fluctuations and large fluctuations in two types of light curves, PDC and SAP, indicating the properties of stellar noise and rotational modulation are highlighted when we apply new MFDMA approach as a function of scale.

The results of this approach indicated in Figure 5 shows average behavior of five multifractal indicators is the slightly same for both pipelines. Meanwhile, the shapes of curves are flat from a maximum period characterized by a rotation period of ~16 days. Our results also demonstrated that the multifractality of Kepler-30 light curves is due to both long-range correlation and broad probability density function, but the long-range correlation is the main source. This result can be emphasized by comparing the original time series with their shuffled and phase-randomized surrogates.

Furthermore, this new multifractal approach can be used to develop theoretical and computational models for various stellar magnetic activity-related phenomena and their interactions with the planets. Finally, these indexes will greatly simplify spot modelling of current TESS and future PLATO data. These last issues will be addressed in a forthcoming communication.

## Acknowledgments

DBdeF acknowledges financial support from the Brazilian agency CNPq-PQ2 (Grant No. 305566/2021-0). Research activities of STELLAR TEAM of Federal University of Ceará are supported by continuous grants from the Brazilian agency CNPq. This paper includes data collected by the *Kepler* mission. Funding for the *Kepler* mission is provided by the NASA Science Mission directorate. All data presented in this paper were obtained from the Mikulski Archive for Space Telescopes (MAST).

## REFERENCES

Basri et al., 2011, ApJ, 141, 20

Basri, G., Walkowicz, L. M., & Reiners, A. 2013, ApJ, 769, 37

Batalha, N. M., Borucki, W. J., Koch, D. G., et al. 2010, ApJL, 713, L109

Belete, A. B., Bravo J. P., Canto Martins B. L., Leão I. C., De Araujo J. M. & De Medeiros J. R., 2018, MNRAS, 478, 3976

Borucki, W., Koch, D., Batalha, N., et al. 2009, IAU Symposium, 253, 289

Borucki W. J. et al., 2010, Science, 327, 977

de Freitas, D. B., & De Medeiros, J. R. 2009, Europhys. Lett, 88, 19001

de Freitas, D. B., Leão, I. C., Lopes, C. E. F., De Medeiros, J. R., et al. 2013, ApJL, 773, L18

de Freitas, D. B., Nepomuceno, M. M. F., de Moraes Junior, P. R. V., Lopes, C. E. F., Leão, I. C. et al. 2016, ApJ, 831, 87

de Freitas, D. B., Nepomuceno, M. M. F., de Moraes Junior, P. R. V., Lopes, C. E. F., Leão, I. C. et al. 2017, ApJ, 843, 103

de Freitas, D. B., Nepomuceno, M. M. F., Alves Rios, L. D., Das Chagas, M. L. & De Medeiros, J. R., 2019a, ApJ, 880, 151

de Freitas, D. B., Nepomuceno, M. M. F., Cordeiro, J. G., Das Chagas, M. L. & De Medeiros, J. R., 2019b, MNRAS, 488, 3274

de Freitas, D. B., Lanza, A. F., Silva Gomes, F. O. & Das Chagas, M. L., 2021, A&A, 650, 40

Fabrycky, D. C., Ford, E. B., Steffen, J. H., et al. 2012, ApJ, 750, 114

Gilliland, R. L., Chaplin, W. J., Jenkins, J. M., Ramsey, L. W., & Smith, J. C. 2015, AJ, 150, 133

Gu, G.-F., & Zhou, W.-X. 2010, Phys. Rev. E, 82, 011136

Hampson, K. M., & Mallen, E. A. H. 2011, Biomedical Optics Express, 2, 464

Hurst, H. E. 1951, Trans. Am. Soc. Civ. Eng., 116, 770

Ihlen, E. A. F. 2012, Front. Physiology 3, 141

Jenkins et al., 2010a, ApJL, 713, L87

Jenkins, J. M., Caldwell, D. A., Chandrasekaran, H., et al. 2010b, ApJL, 713, L120

Kawaler S. D., 1988, ApJ, 333, 236

Koch, D. G., Borucki, W. J., Basri, G., et al. 2010, ApJL, 713, L79

Kraft R. P., 1967, ApJ, 150, 551

Krzyszczak, J.; Baranowski, P.; Zubik, M.; Kazandjiev, V.; Georgieva, V.; Sawinski, C.; Siwek, K.; Kozyra, J.; Nieróbca, A., 2018, Theor. Appl. Climatol., 137, 1811

Lanza, A. F.; Das Chagas, M. L., & De Medeiros, J.R. 2014, A&A, 564, A50

Lanza, A. F., Netto, Y., Bonomo, A. S., et al. 2019, A&A, 626, A38

Makowiec D., & Fuliaski, A. 2010, Acta Phys Pol B, 41, 1025

Maxted, P. F. L. 2016, A&A, 591, A111

Norouzzadeha, P., Dullaertc, W., & Rahmani, B. 2007, Physica A, 380, 333

Panichi F., Gozdziewski K., Migaszewski C., Szuszkiewicz E.,2018, MNRAS, 478, 2480

Sanchis-Ojeda, R., Fabrycky, D. C., Winn, J. N., et al. 2012, Nature, 487, 449

Skumanich A., 1972, ApJ, 171, 565

Tanna, H.J., Pathak, K.N., 2014, Astrophys. Space Sci., 350, 47

Thompson, S. E., Christiansen, J. L., Jenkins, J. M., et al. 2013, Kepler Data Release 23 Notes (KSCI-19063-001)

Walkowicz L. M., & Basri G. S., 2013, MNRAS, 436, 1883

Wang, J., Shang, P., and Cui, X., Phys. Rev. E Stat. Nonlin. Soft Matter Phys. 2014, 89, 032916